\documentclass[letter,aps,nofootinbib,10pt,showpacs,twocolumn]{revtex4-1}

\usepackage{overpic}
\usepackage{slashed}
\usepackage{amsmath}
\usepackage{amssymb}
\usepackage{graphicx} 
\usepackage{hyperref}
\usepackage{tikz}
\usetikzlibrary{arrows}

\usepackage{bm}
\usepackage{bbold}
\usepackage{caption}
\usepackage{subcaption}
\DeclareMathAlphabet{\mathpzc}{OT1}{pzc}{m}{it} 
\usepackage[mathscr]{euscript}
\usepackage{verbatim}

\newcommand{\be}{\begin{eqnarray}}
\newcommand{\ee}{\end{eqnarray}}

\newcommand{\bn}{\begin{enumerate}}
\newcommand{\en}{\end{enumerate}}

\def\Tr{{\rm Tr}}

\def\vec#1{\bm{#1}}

\newcommand{\qed}{\nobreak \ifvmode \relax \else
      \ifdim\lastskip<1.5em \hskip-\lastskip
      \hskip1.5em plus0em minus0.5em \fi \nobreak
      \vrule height0.75em width0.5em depth0.25em\fi}

\allowdisplaybreaks[1]

\usepackage{blindtext}
\usepackage{framed}
\makeatletter
\newcommand\framename{Example}
\newcounter{framecnt}
\newcommand{\TitleFrame}[2]{%
    \fboxrule=\FrameRule
    \fboxsep=\FrameSep
    \vbox{\rlap{\strut#1}\nobreak\nointerlineskip
        \noindent\fbox{#2}}}
\newenvironment{titledframe}[2][\FrameFirst@Lab\ \hspace{-1mm}\emph{(cont.)}]{%
    \refstepcounter{framecnt}%
    \def\FrameFirst@Lab{\emph{\framename\ \theframecnt\ #2}}%
    \def\FrameCont@Lab{\textrm{#1}}%
    \def\FrameCommand##1{%
        \TitleFrame{\FrameFirst@Lab}{##1}}%
    \def\FirstFrameCommand##1{%
        \TitleFrame{\FrameFirst@Lab}{##1}}%
    \def\MidFrameCommand##1{%
        \TitleFrame{\FrameCont@Lab}{##1}}%
    \def\LastFrameCommand##1{%
        \TitleFrame{\FrameCont@Lab}{##1}}%
    \MakeFramed{\hsize0.9\textwidth
    \advance\hsize -2\FrameRule
    \advance\hsize -2\FrameSep
    \FrameRestore}}%
   {\endMakeFramed}
\makeatother

\newcommand{\bframe}{\begin{center}\begin{titledframe}{}\small\flushleft\vspace{-4mm}}
\newcommand{\eframe}{\end{titledframe}\end{center}}

\newcommand{\pal}[1]{{\color{black} #1}}
\begin{document}

\title{Existence and Construction of Galilean invariant $z\neq2$ Theories.}

%

\author{Benjam\'\i{}n Grinstein}
\email{bgrinstein@ucsd.edu}
\affiliation{Department of Physics, 
University of California, San Diego\\
La Jolla, CA 92093, USA}

\author{Sridip Pal}
\email{srpal@ucsd.edu}
\affiliation{Department of Physics, 
University of California, San Diego\\
La Jolla, CA 92093, USA}

\begin{abstract}
  We prove a no-go theorem for the construction of a Galilean boost invariant and $z\neq2$
  anisotropic scale invariant field theory with a finite dimensional basis of fields.  Two point
  correlators in such theories, we show, grow unboundedly with spatial separation.  Correlators of
  theories with an infinite dimensional basis of fields, for example, labeled by a continuous
  parameter, do not necessarily exhibit this bad behavior. Hence, such theories behave effectively
  as if in one extra dimension. Embedding the symmetry algebra into the conformal algebra
  of one higher dimension also reveals the existence of an internal continuous parameter.
  Consideration of isometries shows that 
  the non-relativistic holographic picture assumes a
  canonical form, where the bulk gravitational theory lives in a
  space-time with one extra dimension.
 This can be contrasted with the original proposal by Balasubramanian and McGreevy, and
  by Son, where the metric of a $d+2$ dimensional space-time is proposed to be dual of a $d$
  dimensional field theory.  We 
  provide explicit  examples of theories living at
 fixed point with anisotropic scaling exponent  $z=\frac{2\ell}{\ell+1}\,,\ell\in \mathbb{Z}$.
\end{abstract}

	

\maketitle

\section{Introduction}
Gravity duals of non-relativistic field theories have been proposed in
\cite{Balasubramanian:2008dm,Son:2008ye}. It has been observed in
Ref.~\cite{Balasubramanian:2008dm}, that one can consistently define
an algebra with Galilean boost invariance and arbitrary anisotropic
scaling exponent $z$. While the metric having isometry of this
generalized Schr\"odinger group has been used with the holographic
dictionary to construct correlators of a putative field
theory\cite{Ross:2009ar,Guica:2010sw,Costa:2010cn, Guica:2011ia,
  vanRees:2012cw,Andrade:2014iia,Andrade:2014kba}, there is no
explicit field theoretic realization of such a symmetry for
$z\neq2$.\footnote{
  We note that just matching the isometries is necessary but not
  sufficient for the existence of a holographic description. Here we
  just seek a group invariant field theory, which may or may not have
  a gravity dual.} One surprising feature, noted as a ``strange
  aspect" in Ref.~\cite{Balasubramanian:2008dm}, is that, unlike in
the canonical AdS/CFT correspondence, where the CFT in $d$ dimensions
is dual to the gravity in $d+1$-dimensions, in the non-relativistic
case the metric is of a space-time with two additional dimensions. The
$(d+2)$-dimensional metric, having isometries of the $d$-dimensional
generalized Schr\"odinger group, is given
by\cite{Balasubramanian:2008dm,Son:2008ye}
\begin{align}
ds^2=L^2\left[-\frac{dt^2}{r^{2z}}+\frac{2d\xi dt+dx^2}{r^2}+\frac{dr^2}{r^2}\right],
\end{align}
where $\xi$ is the extra dimension having no analogous appearance in
the relativistic AdS-CFT correspondence. The metric is invariant under
the required anisotropic scaling symmetry
\begin{align}
\label{eq:dilatations}
x_i \to \lambda x_i\,,\ t \to \lambda^z t\,,\ r\to \lambda r\,,\ \xi \to \lambda^{2-z} \xi,
\end{align}
and under Galilean boosts
\begin{align}
x_i&\to x_i + v_it\,, \qquad \xi \to \xi - \frac{1}{2} \left(2v_ix_i+v^2t\right).
\end{align}
For $z=2$, an explicit construction of Galilean boost invariant field
theory in $(d-1) +1$ dimensions has been known. Thus a question arises
naturally as to whether one can get rid of the extra $\xi$ direction
and reduce the correspondence down to a canonical correspondence
between a $d$-dimensional quantum field theory on flat space and a
$(d+1)$-dimensional gravitational theory. This was answered positively
in Ref.~\cite{Balasubramanian:2010uw}. But for $z\neq2$ we do not know
of any explicit $d$-dimensional field theoretic example having the
generalized Schr\"odinger symmetry, nor do we know an example of a
$(d+1)$-dimensional metric having the same set of isometries. Thus the
 ``strange aspect" of $d$-$(d+2)$ correspondence appears to persist
for $z\neq2$.

In this paper, we initiate a field theoretic study of $z\neq2$
theories.\footnote{Theories with $z=2$ have been studied from a field
  theoretic point of view in many works; see, {\it e.g.},
  Refs.~\cite{Nishida:2007pj,Nishida:2010tm,Goldberger:2014hca,Golkar:2014mwa}. The
  $z=\infty$ case without particle number symmetry has been explored in
  Ref.~\cite{Hofman:2014loa,Jensen:2017tnb}.}   We prove a no-go theorem for the construction of a space-time translation invariant, 
  rotation invariant, Galilean boost invariant,\footnote{Here by Galilean boost invariance, we mean invariance under both the boost and a $U(1)$ particle number symmetries. The $U(1)$ naturally arises as a commutator of generators of boosts and translations.} and $z\neq2$
  anisotropic scale invariant field theory with a finite number\footnote{More precisely, a
    finite-dimensional basis of operators as defined below Eq.~\eqref{genK}} of fields in $d$
  dimensions. Two point correlators in such theories, we show, grow unboundedly with spatial separation.  By
 contrast, correlators of theories with an infinite number of fields, {\it e.g.}, labeled by a continuous parameter, do not necessarily exhibit this bad behavior.  Hence, such theories behave effectively as a $(d+1)$-dimensional theory. In the context of holography, this
explains the  ``strange aspect"; the $z\neq2$ theories indeed
provide us with the possibility of a canonical realization of
holography, {\it i.e.}, a $(d+1)$-dimensional theory is dual to a
$(d+2)$-dimensional geometry. The $z=2$ case is special in that respect since it is
possible to obtain a $d$-dimensional theory with a finite number of fields such that the symmetries
on field theory side match onto the isometries of a $(d+2)$-dimensional geometry. The special role
of $z=2$ has been emphasized in the context of the 
holographic dictionary in Refs.~\cite{Andrade:2014iia,Andrade:2014kba}.  For $z=2$, the dual
space-time can be made into a $(d+1)$-dimensional one via Kaluza-Klein
reduction of the $(d+2)$-dimensional metric
\cite{Balasubramanian:2010uw}. This is possible since,  for $z=2$,  the extra direction $\xi$ does
not scale by the transformations given in Eq.~\eqref{eq:dilatations}. The scaling of  $\xi$ given in
Eq.~\eqref{eq:dilatations}  can be  verified on the field
theory side of the duality by embedding the $d$-dimensional generalized Schr\"odinger
group into the conformal group of one higher dimension
{\it i.e.}, $SO(d,2)$. By contrast, since for $z\neq2$ the $\xi$ direction does scales,  any
attempt to compactify the extra direction $\xi$  is at odds with the continuous scaling
symmetry. The no-go theorem that we have proved is consistent with the argument in Ref.~\cite{deBoer:2017ing}, based on
 consistency of thermodynamic equation of state, that a perfect fluid with $z\neq2$ Schr\"odinger
 symmetry and discrete spectrum for the energy and particle number, $H$ and $N$, can not exist. In Sec.~\ref{sec:examples}, we present some 
fixed point theories  with $z=\frac{2\ell}{\ell+1}$, with $\ell\in\mathbb{Z}$.

Before delving into a technical proof,  we present a physical argument for our main
  result.\footnote{We thank John McGreevy for discussions leading to this argument.} Consider a
  theory invariant under $z=2$ Schr\"odinger symmetry, where, under a boost\cite{Balasubramanian:2008dm}
\begin{align}
\phi(\vec{x},t)\mapsto \exp\left[-\imath n(\tfrac{1}{2}v^2t+\vec{v}\cdot\vec{x})\right]\phi(\vec{x}-\vec{v}t,t)\,,
\end{align}
where $[N,\phi]=n\phi$. In turn, the state of a particle with momentum $\vec{k}=0$ {\it i.e.},
$\phi^\dagger_{\vec{k}=0}|0\rangle$  transforms under the boost by $\vec{v}$ as follows:
\begin{align}
\nonumber |\vec{v}\rangle &\equiv e^{-\imath \vec{K}\cdot\vec{v}} \phi^\dagger_{\vec{k}=0}|0\rangle \\
\nonumber &= \int\!\! d\vec{x}\ \exp\left[\imath n(\tfrac{1}{2}v^2t+\vec{v}\cdot\vec{x})\right] \phi^\dagger(\vec{x}-\vec{v}t,t|0\rangle\\
\label{boostedparticle}
&=\exp\left[\imath \tfrac{nv^2}{2}t\right] \phi^\dagger_{\vec{k}=n\vec{v}}|0\rangle\,.
\end{align}
This  has the interpretation of having a boosted particle moving with momentum $n\vec{v}$ and
kinetic energy $-\tfrac{1}{2} n\vec{v}^2$.  A positive value of $n$ results in decreasing energy with
increasing boost.  Therefore,   negative semi-definiteness of $n$ is required for  stability. In
case of  more than a single species of particle, the matrix $\mathbb{N}$ appearing in
$[N,\Phi^\dagger]=-\mathbb{N}\Phi^\dagger$ has to be negative semi-definite.  As we will see, from the
symmetry algebra it follows  that for a
theory with finite number of fields with $z\neq2$ the trace of $\mathbb{N}$ must vanish,  spoiling the
negative semi-definiteness and the stability in the sense discussed above; by contrast, for $z=2$
there is no constraint on the trace of $\mathbb{N}$. The above is merely a
heuristic argument, giving  intuition behind the technical result presented below.

\section{Generalized Schr\"odinger algebra and its Representation} 
\label{main}
The Galilean algebra consists of generators corresponding to spatial
translations, $P_i$, time translation, $H$, Galilean boosts, $K_i$,
rotations, $M_{ij}$, along with a particle number generator, $N$, such that they satisfy the following commutation relations \cite{Nishida:2007pj,Nishida:2010tm,Goldberger:2014hca,Golkar:2014mwa}:
\begin{gather}
[M_{ij},N]=[P_i,N]=[K_i,N]=[H,N]=0\nonumber \\
 [M_{ij}, P_{k}]=\imath
(\delta_{ik}P_{j}-\delta_{jk}P_{i})\,, \nonumber \\
 [M_{ij},K_{k}]=\imath
(\delta_{ik}K_{j}-\delta_{jk}K_{i})\,, \nonumber\\
[M_{ij},M_{kl}]= \imath (\delta_{ik}M_{jl}-\delta_{jk}M_{il}+\delta_{il}M_{kj}-\delta_{jl}M_{ki})\,,\nonumber \\
\label{eq:GalAlg1} 
[P_{i},P_{j}]=[K_{i},K_{j}]=0\,, \qquad [K_{i},P_{j}]=\imath\delta_{ij}N\,,\\
[H,N]=[H,P_{i}]=[H,M_{ij}]=0\,, \quad [H,K_{i}]=-\imath P_{i}\,. \nonumber  
\end{gather}
The algebra can be enhanced  by appending a dilatation generator $D$,\footnote{\pal{This enhanced
    algebra corresponds to that of deformed ISIM(2) \cite{Gibbons:2007iu}, with the following identification: $H\mapsto P_+$, $N\mapsto P_-$, $K_i\mapsto M_{+i}$ and $D\mapsto -\tfrac{1}{b}N$ where $b(z-1)=1$.}}
which scales space and time separately, in the following way:
\begin{align}
x_i \to \lambda x_i,& \qquad t \to \lambda^z t\,.
\end{align}
The commutators of $D$ with the rest of the generators are given by
\begin{gather}
[D,P_{i}]=\imath P_{i}\,, \quad [D,K_{i}]= (1-z)\imath K_{i}\,,  
\quad [D,H]=z\imath H\,,\nonumber \\
\label{eq:GalAlg2} 
\quad [D,N]=\imath (2-z)N\,, \quad [M_{ij},D]=0\,.
\end{gather}

The physical interpretation of $N$ is subtle. For $z=2$ it is usually
thought of as a particle number symmetry generator. The subtlety in
the context of holography has been explored in
\cite{Balasubramanian:2010uw}. For rest of this work, we take an
agnostic viewpoint and treat $N$ as a generator of symmetry without
specifying its physical origin. This will enable us to explore all the
possibilities, as allowed by symmetries. The case $z=2$ is very
special in that one can append an additional generator $C$ of special
conformal transformations. Thus one can have the full Schr\"odinger
algebra for $z=2$ \cite{Hagen:1972pd,Niederer:1972zz,Mehen:1999nd,Henkel:2003pu,Goldberger:2014hca,Golkar:2014mwa,Pal:2018idc}. When $z\neq 2$, the generator corresponding to
special conformal transformation is not available.

In what follows, we will assume (unless otherwise specified) that the
field theory lives in $d=(d-1)+1$ dimensions and that the vacuum is invariant under Galilean boosts,  {\it i.e.}, $K_i|0\rangle =\langle 0|K_i =0$

The field representation is built by defining local operators
$\Phi$ such that $H$ and $P$ act canonically,
\begin{align}
\label{eq:canonicalHP}
[H,\Phi]=-\imath\partial_t\Phi\,,\ [P_i,\Phi]= \imath\partial_i\Phi\,.
\end{align}
%
%
%
We consider representations of the little group, generated by $D$, $K_i$, $N$ and
  $M_{ij}$, that keeps the origin, $(\vec{0},0)$, invariant. The fields $\Phi$  have definite
  transformation properties under $D$, $K_i$ and $N$,\footnote{The fields $\Phi$  also have definite
    transformation properties under $M_{ij}$, but this will not play a role in the discussion below.}
\begin{align}\label{genD}
\left[D,\Phi(\vec{x}=\vec{0},t=0)\right]&=\imath\mathcal{D}\Phi(\vec{x}=\vec{0},t=0)\,,\\
\left[N,\Phi(\vec{x}=\vec{0},t=0)\right]&=\mathcal{N}\Phi(\vec{x}=\vec{0},t=0)\,,\\
\label{genK}\left[K_i,\Phi(\vec{x}=\vec{0},t=0)\right]&=\mathcal{K}_i\Phi(\vec{x}=\vec{0},t=0)\,.
\end{align}
where $\mathcal{D}$, $\mathcal{N}$, and $\mathcal{K}_i$ are linear operators. We refer to
  the smallest non-trivial irreducible representation in Eqs.~\eqref{genD}--\eqref{genK} as ``the
  basis of operators".\footnote{For example, the free Schr\"odinger field theory is invariant under
    $z=2$ Schr\"odinger algebra and the single field $\phi$ forms a one dimensional irreducible
    representation of the little group i.e. $\left[D,\phi(\vec{0},0)\right]=\imath
    \tfrac{d}{2}\phi(\vec{0},0)\,,\ \left[N,\phi(\vec{0},0)\right]=N\phi(\vec{0},0)$ and
    $\left[K_i,\phi(\vec{0},0)\right]=0$.} For Lagrangian theories the basis of operators
  corresponds to the elementary fields from which the Lagrangian is constructed.
Henceforth, we restrict our attention to the basis of operators, and continue to denote by
$\mathcal{D}$, $\mathcal{N}$, and $\mathcal{K}_i$ their linear representation. 
 In the finite dimensional case, we denote these by finite dimensional matrices $\mathbf{\Delta}$, $\mathbb{N}$ and $\mathbb{K}_i$,
respectively.

Consider $G_{\alpha\beta}\equiv \langle 0|\Phi_{\alpha}(x,t)\Phi_{\beta}(0,0)|0\rangle$. Using
Eqs.~\eqref{eq:canonicalHP}, the commutator in~\eqref{genK}
translates, in the finite dimensional case, to
\begin{align}
[K_i, \Phi]&= \left(-\imath t\partial_{i} \mathbf{I}+ x_i\mathbb{N}+\mathbb{K}_i\right)\Phi
\end{align}
where $x_{i}=x^{i}$. Galilean boost invariance of the vacuum, $K_i|0\rangle =\langle 0|K_i=0$, then
gives
\begin{multline}
\nonumber \langle 0|\left[K_i,\Phi_\alpha(x,t)\Phi_\beta(0,0)\right]|0\rangle = 0\\
\Rightarrow \left(-\imath t\partial_{i} \delta_{\alpha\sigma}+ x_i\mathbb{N}_{\alpha\sigma}+\mathbb{K}_{i\alpha\sigma}\right)G_{\sigma\beta}+\mathbb{K}_{i\beta\sigma}G_{\alpha\sigma}=0\,.
\end{multline}
Using the fact that  $[\mathbb{N},\mathbb{K}_i]=0$, the solution to the above differential equation  is given by 
\begin{equation}
\nonumber G_{\alpha\beta}=
\bigg(\!\!
e^{-\imath\frac{|\vec{x}|^2}{2t}\mathbb{N}}e^{-\imath\frac{\vec{x}\cdot\vec{\mathbb{K}}}{t}} C(t) e^{-\imath\frac{\vec{x}\cdot\vec{\mathbb{K}^{T}}}{t}} \bigg)_{\alpha\beta}
\end{equation}
where $C(t)$ is an as yet undetermined matrix function of $t$. The norm is defines as $|\vec{x}|^2\equiv\sum_i(x^{i})^2$ while the dot product is defined as $\vec{x}\cdot\vec{K}=\sum_ix_iK_i$. Similarly, one can consider $G^\prime_{\alpha\beta}\equiv \langle 0|\Phi_{\alpha}(x,t)\Phi^\dagger_{\beta}(0,0)|0\rangle$ which is given by
\begin{align}
\nonumber G^\prime_{\alpha\beta}=
\bigg(\!\!
  e^{-\imath\frac{|\vec{x}|^2}{2t}\mathbb{N}}e^{-\imath\frac{\vec{x}\cdot\vec{\mathbb{K}}}{t}} C^\prime(t) e^{\imath\frac{\vec{x}\cdot\vec{\mathbb{K}^{\dagger}}}{t}} \bigg)_{\alpha\beta}
\end{align}
where $C^\prime(t)$ is  as yet undetermined.

Since $[D,N]=\imath (2-z)N$, we have
\begin{equation*}
[\mathbf{\Delta},\mathbb{N}]=(2-z)\mathbb{N}\,.
\end{equation*}
and this leads to $\Tr(\mathbb{N})=0$; similarly, for $z\ne1$ we have $\Tr(\mathbb{K}_i)=0$. Now using Jordan-Chevalley decomposition, we can write
\begin{align*}
\mathbb{N}&=\mathbb{N}_1+\mathbb{N}_2\,,\ \left[\mathbb{N}_1,\mathbb{N}_2\right]=0\,,\\
\mathbb{K}&=\mathbb{K}_1+\mathbb{K}_2\,,\ \left[\mathbb{K}_1,\mathbb{K}_2\right]=0\,.
\end{align*}
where $\mathbb{K}_1$ and $\mathbb{N}_1$ are diagonalizable matrices while
$\mathbb{N}_2$ and $\mathbb{K}_2$ are nilpotent matrices (here and
below we 
suppress the vector index in $\mathbb{K}$ and $\mathbb{K}_{1,2}$ to
avoid clutter). Let us define diagonal matrices $D_{N_1}$ and $D_{K_1}$ such that 
\begin{align}
P_ND_{N_1}P^{-1}_N=\mathbb{N}_1,\quad P_KD_{K_1}P^{-1}_K=\mathbb{K}_1
\end{align}
where $P_N$ and $P_K$ diagonalize $\mathbb{N}_1$ and $\mathbb{K}_1$ respectively.
The zero trace condition leads to $\Tr(\mathbb{K}_1)=\Tr(D_{K_1})=0$ and
$\Tr(\mathbb{N}_1)=\Tr(D_{N_1})=0$, which in turn implies that either all the
diagonal entries of $D_{N_1}$ (or $D_{K_1}$) are zero, in
which case $\mathbb{N}$ (or $\mathbb{K}$) is a nilpotent matrix,  or
there has to be both positive and negative entries. We can then recast
the correlators as follows:
\begin{align}\label{maincorrelator}
G_{\alpha\beta}&=
\bigg(\!\!
                 e^{\frac{|\vec{x}|^2}{2\imath t}\mathbb{N}_1}e^{\frac{|\vec{x}|^2}{2\imath t}\mathbb{N}_2}e^{\frac{\vec{x}\cdot\vec{\mathbb{K}_1}}{\imath t}} e^{\frac{\vec{x}\cdot\vec{\mathbb{K}_2}}{\imath t}}
C(t) 
e^{\frac{\vec{x}\cdot\vec{\mathbb{K}^{T}_1}}{\imath t}} e^{\frac{\vec{x}\cdot\vec{\mathbb{K}^{T}_1}}{\imath t}}\bigg)_{\alpha\beta}\\
G^\prime_{\alpha\beta}&=
\bigg(\!\!
                        e^{\frac{|\vec{x}|^2}{2\imath t}\mathbb{N}_1}e^{\frac{|\vec{x}|^2}{2\imath t}\mathbb{N}_2} e^{\frac{\vec{x}\cdot\vec{\mathbb{K}_1}}{\imath t}} e^{\frac{\vec{x}\cdot\vec{\mathbb{K}_2}}{\imath t}} 
C^\prime(t) 
e^{-\frac{\vec{x}\cdot\vec{\mathbb{K}^{\dagger}_{1}}}{\imath t}}e^{-\frac{\vec{x}\cdot\vec{\mathbb{K}^{\dagger}_{2}}}{\imath t}} \bigg)_{\alpha\beta}
\end{align}

It follows that when $\mathbb{N}_1\neq 0$,
$e^{\frac{|\vec{x}|^2}{2\imath t}\mathbb{N}_1}=P_Ne^{\frac{|\vec{x}|^2}{2\imath t}D_{N_1}}P_N^{-1}$ has exponential growth
for imaginary time irrespective of how we do the analytical
continuation of the correlator to imaginary time. This growth can not be overcome by
any of the other terms as nilpotency of $\mathbb{N}_2$ guarantees that
\begin{align}
e^{-\imath\frac{|\vec{x}|^2}{2t}\mathbb{N}_2}=\sum_{\ell=0}^{\ell=M-1}\left(-\imath\frac{|\vec{x}|^2}{2t}\right)^{\ell}\mathbb{N}_{2}^{\ell}
\end{align}
where $\mathbb{N}_{2}^{M}=0$ for some integer $M$. Also, terms like
$e^{\imath\frac{\vec{x}\cdot\vec{\mathbb{K}}}{t}}$ cannot suppress the
exponential growth arising from $\mathbb{N}_1$.

If instead $\mathbb{N}_{1}=0$ then
$e^{-\imath\frac{|\vec{x}|^2}{2t}\mathbb{N}_2}$ gives polynomial growth
with $x$. We employ the same technique to establish the effect of
$e^{\imath\frac{\vec{x}\cdot\vec{\mathbb{K}}}{t}}$. If
$\mathbb{K}_{1}\neq 0$, there will be exponential growth for some
entries, while terms involving $\mathbb{K}_{2}$ are polynomial in
nature, giving exponential growth as a whole. Alternatively, if
$\mathbb{K}_{1}=0$ then $\mathbb{K}$ is nilpotent and we have polynomial
growth.

We note that only when $z=2$ or the representation is infinite, we can
not implement the $\Tr(\mathbb{N})=0$ condition and the above argument
fails. This is expected for $z=2$ since the two point correlator is
well behaved in this case, that corresponds to Schr\"odinger field
theory \cite{Hagen:1972pd,Niederer:1972zz,Mehen:1999nd,Henkel:2003pu}. We conclude that in the finite dimensional case for $z\ne 2$ a
quantum field theory with the symmetry of the algebra in
Eqs.~\eqref{eq:GalAlg1} and~\eqref{eq:GalAlg2} is ill-behaved. For example, since correlators grow
with spatial separation  cluster decomposition  fails. The same conclusion can be drawn via
an independent argument in the case that $\mathbf{\Delta}$ is diagonal; see App.~\ref{diagonalizableD}.

Therefore, for $z\ne2$ we are left to consider infinite dimensional
representations. In this case we can display explicitly an example
that does not obviously lead to problematic quantum field theories.
To achieve this, we introduce fields $\psi$ labeled by a new
non-compact variable~$\xi$, such that
\begin{align}
\left[N,\psi\right]&=\imath\partial_{\xi}\psi\,,\\
\label{nondiagonalD}
\left[D,\psi\right]&=\imath
                     \left(zt\partial_t+x^i\partial_i+(2-z)\xi\partial_{\xi}+\Delta_{\psi}\right)\psi\,,\\
\label{primaryK}
\left[K_i,\psi\right]&= \left(-\imath t\partial_{i}+ \imath x_i \partial_{\xi}\right)\psi.
\end{align}
Thus, $\mathcal{D}=(2-z)\xi\partial_{\xi}+\Delta_{\psi}$,
$\mathcal{N}=\imath\partial_{\xi}$, and
$\mathcal{K}_i= 0$. Note that $\xi$ must be a
non-compact variable, else scaling symmetry is broken. To be concrete,
\begin{align}
\left[\xi,\partial_{\xi}\right]=-1\,,\quad \left[\xi\partial_{\xi},\partial_\xi\right]=-\partial_\xi\,,
\end{align}
are well defined only when $\xi$ is a non-compact variable. If we take a Fourier transform with
respect to $\xi$, it becomes obvious that $\mathcal{N}$ is diagonal while $\mathcal{D}$ is not
diagonal. This, however, is immaterial, since in terms of a new variable $\xi'=\ln|\xi|$, 
$\mathcal{N}$ is non-diagonal and $\mathcal{D}$ is diagonal.

We say $\psi$ is a {\it primary} operator if
$[K_i,\psi(\vec{x}=0,t=0;\xi)]=0$, that is, $\mathcal{K}_i=0$; this
was assumed in the commutation relations \eqref{primaryK}.  Once
again, one can invoke the Galilean boost invariance of the vacuum to
obtain the form of the two point correlator of primaries $\psi$ and
$\phi$. This is most easily computed in terms of the the Fourier
transformed operators,  {\it e.g.}, $\psi(\vec{x},t,m_1)=\int d\xi\
\psi(\vec{x},t,\xi)e^{\imath m \xi}$; we obtain
\begin{align}
&\langle 0|\psi(\vec{x},t,m_1)\phi(0,0,m_2)|0\rangle 
\nonumber\\
 &=
\begin{cases}
  h(t)\delta\left(m_1+m_2\right)f(t^{2-z}m_1^{z})\exp\left(\frac{\imath
      m_{1}|\vec{x}|^2}{2t}\right)\,,  &z\neq 0\\
  \label{sch}
h(t)\delta\left(m_1+m_2\right)f(m_1)\exp\left(\frac{\imath
    m_{1}|\vec{x}|^2}{2t}\right)\,, &z= 0
\end{cases}
\end{align}
where $h(t)$ is an as yet undetermined function of $t$. Evidently,
Eq.~\eqref{sch} is consistent with the correlator of the $z=2$ theory
\cite{Goldberger:2014hca,Golkar:2014mwa}. For $z\neq2$, rewriting in
terms of $\xi$, we obtain:
\begin{align}
\nonumber
 \langle &0|\psi(\vec{x},t,\xi)\phi(0,0,0)|0\rangle\\
\label{corr}&\propto 
\begin{cases}
h(t)t^{1-2/z}
\tilde{g}\left(\frac{|\vec{x}|^2-2t\xi}{2t^{2/z}}\right)\,, & z\neq0\\
 \tilde{h}(t)\tilde{f}\left(\frac{|\vec{x}|^2}{2t}-\xi\right)=\tilde h(t)\left(\frac{|\vec{x}|^2}{2t}-\xi\right)^{-\Delta/2}\,,
 &z=0
 \end{cases}
\end{align}
where $\tilde{g}(s)=\int dy\ e^{-\imath y s}g(y)$, $g(y)=f(y^{z})$ and
$y^{z}=m^zt^{2-z}$. When $z=0$, we use the fact that $\tilde{f}$ has
to scale covariantly under $z=0$ scaling, where $\tilde{f}$ is the Fourier
transform of $f$; here $h(t)$ must be a power law of $t$ with $t^{-\alpha}$ such that the scaling dimensions of $\psi$ and $\phi$ add
up to $\alpha z+(2-z)$ for $z\neq0$ and $\Delta$ for $z=0$ with $\tilde{h}(t)$ being any function of $t$.

\section{Null reduction and Embedding into Conformal group $SO(d,2)$}\label{embed}
A standard trick to obtain a $d$ dimensional $z=2$ Schr\"odinger
invariant theory is to start with a conformal field theory in $d+1$
dimensions and perform a null cone reduction \cite{Maldacena:2008wh,Jensen:2014hqa,Banerjee:2015uta,Banerjee:2015hra,Pal:2017ntk,Grosvenor:2017dfs}. This is possible because the
Schr\"odinger group, $\text{Sch}(d)$, can be embedded into
$SO(d,2)$. Next we show that the generalized Schr\"odinger group
can also be embedded into $SO(d,2)$. {\pal{A similar embedding has been considered in
    \cite{Gibbons:2007iu} in the context of the Lie algebra of the deformed ISIM(2) group.}}

If the generators of $SO(d,2)$ are given by $P^{(r)}_{\mu}, M^{(r)}_{\mu\nu}, D^{(r)}, C^{(r)}_{\mu}$ where $P^{(r)}_{\mu}$ are translation generators, $M^{(r)}_{\mu\nu}$ are Lorentz generators, $D^{(r)}$ is the relativistic scaling generator and $C^{(r)}_{\mu}$ are special conformal generators (here the superscript ``(r)" denotes the relativistic generators), then following generators generate the generalized Schr\"odinger algebra: 
\begin{align}
K_{i}&=M^{(r)}_{i-}\,,H=P^{(r)}_{+}\,, N=P^{(r)}_{-}\\
M_{ij}&=M^{(r)}_{ij}\,,P_i=P^{(r)}_i\\
D&=D^{(r)}+(1-z)M^{(r)}_{+-}
\end{align}
It is straightforward to verify that $D$ scales
$x^{-}\to \lambda^{2-z}x^{-}$. Only for $z=2$, does $x^{-}$ not scale
and one is able to do a null cone reduction via compactification in
the $x^{-}$ direction, yielding a discrete spectra for $N$. On the
other hand, for $z\neq2$, even via null cone reduction one can not
truly get rid of the $x^{-}$ direction since any compactification in
the $x^{-}$ direction would spoil the scaling symmetry. As a result,
for $z\neq2$ the null reduction always leaves a continuous spectra for
the generator $N$.

\section{Explicit $d+1$ dimensional examples}
\label{sec:examples}
\subsection{$z=0$}
Here we provide with an explicit example of a generalized Schr\"odinger
invariant theory in $(d-1)+1$ dimensions with  $z=0$  and verify that the two point correlator
indeed conforms to the general form given in
Eq.~\eqref{corr}. 

We consider a Lagrangian model given by  
\begin{align}\label{z0}
\mathcal{L}=\phi^{\dagger}\left(2\partial_t\partial_\xi-\nabla^2+2\imath \partial_{\xi}\right)\phi
\end{align}
and the two point correlator is given by\footnote{\pal{The correlator in \eqref{z0twopoint} follows
  from~\eqref{z0} only after restricting the field $\phi$ to positive $\xi$-Fourier modes; see the
  footnote below Eq.~\eqref{eq:corrxtm} for more details.}}
\begin{align}\label{z0twopoint}
\langle \phi \phi^{\dagger}\rangle \propto \left(\frac{1}{t}\right)^{\tfrac{d-1}{2}}\exp\left[-\imath t\right] \left(\frac{|\vec{x}|^{2}}{2t}-\xi\right)^{-\tfrac{d-1}{2}}
\end{align}
In $d+1$ dimensions, $\frac{d-1}{2}=\frac{(d+1)-2}{2}$ is precisely the dimension of a free
  relativistic scalar. This is because the  generalized Schr\"odinger algebra can be embedded into
  the conformal group of one higher dimension, as mentioned in Sec.~\ref{embed}.
 
For $z=0$, $t$ does not scale. One may contemplate perturbing the gaussian fixed point by 
   marginal operators constructed out of powers of $\partial_{t}$, for example, $\phi^\dagger
   \exp\left(\imath\partial_t\right)\partial_\xi\phi$. However, Galilean boost invariance requires
   that $\partial_t$ appears in the combination with other  derivatives shown in Eq.~\eqref{z0}.   By
   contrast, in  the models presented in Refs.~\cite{Hofman:2014loa,Jensen:2017tnb}, where $N=0$ and the
   Lagrangian is invariant under $\vec{x}\to\vec{x}$ and $t\to\lambda t$, arbitrary powers of spatial
   derivatives are allowed. 
 
\subsection{$z=\frac{2\ell}{\ell+1}\,, \ell \in \mathbf{Z}$, $\ell\geq 1$}
These series of examples are given by following Lagrangian
\begin{align}\label{ell}
\mathcal{L}_{\ell}= \phi^{\dagger}\left(2\partial_t\partial_\xi-\nabla^2+2g\left(\imath \partial_{\xi}\right)^{\ell+1}\right)\phi
\end{align}
The two point correlators, after partial Fourier transformation is given by
\begin{align}
\label{eq:corrxtm}
G(\vec{x},t,m) \propto t^{-\tfrac{d-1}{2}} m^{\tfrac{d-3}{2}}\exp\left[\imath\left(\frac{m|\vec{x}|^2}{2t}-gm^{\ell}t\right)\right]
\end{align}
where $z=\frac{2\ell}{\ell+1}$. One can Fourier transform\footnote{{Care is needed regarding the
    allowed values of $m$. The correlator in \eqref{eq:corrxtm} is most readily obtained by Fourier
    transform of 
\begin{align}
G(\vec{k},t,m) \propto \exp\left[-\imath t\left(\frac{|\vec{k}|^2}{2m}+gm^\ell\right)\right]\,.
\end{align}
For even, positive $\ell$, the integral over  $\vec{k}$ is well defined only for $Im(t/m)<0$, and the result can
be analytically continued to  all values of $t/m$. The integral over $m$ requires $Im(t)<0$ (for
$g>0$), and again one analytically continues to all values of $t$. 

For odd (and positive) $\ell$, the Fourier transform with respect to $m$ is ill behaved for any
value of $t$, because there is no deformation of the contour of integration that can render the
integral of $\exp\left[\imath\left(\frac{m|\vec{x}|^2}{2t}-gm^{\ell}t\right)\right]$ over $m$ finite. Both
for $\ell$ odd and for $\ell=0$, a sensible way to make this integral well defined is to restrict it
to $m>0$. This is, in fact, how we obtained the correlator for the $z=\ell=0$ in
Eq.~\eqref{z0twopoint}. Strictly speaking, these are not Lagrangian theories; these systems are
close analogues of the chiral boson, where the Fourier modes are
restricted~\cite{imbimbo1987lagrangian}.}}  to obtain the correlator in position space-time only
depending on the analytical ease to do so. For $d=3$, $\ell=2$ i.e $z=\frac{4}{3}$, we have
\begin{align}
G(\vec{x},t,\xi) \propto t^{-1} \frac{1}{\sqrt{gt}}\exp\left[\frac{\imath(x^2-2\xi t)^2}{16gt^3}\right]
\end{align}
which is consistent with Eq.~\eqref{corr} for $z\neq0$. \pal{After performing a Euclidean rotation, $t\to
-i\tau$, $\xi\to i\xi$,  one finds good behavior of this correlator at large spatial separation (as
long as $g<0$). }

One can add classically marginal interactions to the model in \eqref{ell}. For example, one may  add
$(\phi\phi^\dagger)^{n-1}\phi(\imath\partial_{\xi})^{k}\phi^\dagger$ with
$k=(\ell+1)[(d-1)\beta+d-2]$ and  $n=2\beta+3$, where  $\beta$ is a non-negative integer. Furthermore, one can have supersymmetric generalizations of $z\neq2$ theories, much like  the $z=2$ case presented in \cite{Meyer:2017zfg} where supersymmetry is an internal symmetry exchanging Fermionic and Bosonic fields.

\section{Conclusion}
The most natural way to realize the Schr\"odinger algebra and its $z\neq 2$ avatar in a gravity dual
of a $d$-dimensional non-relativistic  field theory with Galilean boost and scale invariance with
dynamical exponent $z$ is via isometries of the bulk metric. As it turns out,  the dual metric is of
a $(d+2)$-dimensional space-time \cite{Balasubramanian:2008dm,Son:2008ye}. By contrast, for the canonical
notion of gauge-gravity duality the bulk gravitational theory lives in one extra dimensional
space-time. Above we have expounded the presence of the two extra dimensions in the duality. We
showed that on the field theory side of the duality, for $z\neq 2$, one needs to have an internal
continuous parameter, effectively making the field theory $(d+1)$-dimensional. Any attempt to
construct a $z\neq2$ non-relativistic  field theory with Galilean boost and scale invariance with
finite number of fields  is bound to run into trouble, since correlators will grow with separation
and  will fail to exhibit  cluster decomposition. This result follows solely  from constrains that
the symmetry algebra places on two point correlators. It is important to have the particle number symmetry for the no-go theorem. Without particle number symmetry, there are indeed examples of Galilean boost
invariant $z\neq2$ theories \cite{Stichel:2010ue}. Examples of theories with $z=\infty$ anisotropic scaling symmetry based on warped conformal field theories, are discussed in Ref.~\cite{Hofman:2014loa,Jensen:2017tnb}.

Only for $z=2$ is a consistent $d$-dimensional field theoretic realization of the symmetry, with
finite number of fields, possible, and therefore a conventional $(d+1)$-dimensional gravity dual is
available.  On the gravity side, the metric dual to a $z=2$ Schr\"odinger theory has a direction
$\xi$ which does not scale, and can therefore be compactified.  The 
Kaluza-Klein reduction of the momentum conjugate to $\xi$ generates a discrete spectrum for $N$ that
matches onto a $d$-dimensional field theory. The $\xi$ direction for $z\neq 2$ duals scales,
 forbidding any such compactification. One can also see this by
embedding the generalized Schr\"odinger group into $SO(d,2)$; see Sec.~\ref{embed}. 

That there is no impediment to constructing a sensible $z\ne2$ non-relativistic 
field theory with Galilean boost and scale invariance for an infinite number of fields is most
easily established by giving explicit examples.   Above we presented  explicit
 examples of Galilean boost invariant theories, with $z=\frac{2\ell}{\ell+1}$.


Given that we have explicit examples and the generic form of the correlator, several new
 questions come to mind. One can ask how one may couple these theories to gravity. Non-relativistic
 theory coupled to gravity gives a natural framework to study Ward identity anomalies, and  scale
 anomalies~\cite{Banerjee:2014pya,Banerjee:2014nja,Jensen:2014aia,Geracie:2014nka,Son:2013rqa,Baggio:2011ha,Jensen:2014hqa,Arav:2014goa,Arav:2016xjc,Pal:2016rpz,Auzzi:2016lrq,Pal:2017ntk}. Since these theories are intrinsically $(d+1)$-dimensional, the use of Newton-Cartan geometry is not a natural choice.
It would also be interesting to understand the dispersion relation of Goldstone bosons, arising from
spontaneous breaking of $z\neq2$ scale symmetry; the $z=2$ case has been studied in \cite{Arav:2017plg}.

\section*{Acknowdgement}
The authors thank Kristan Jensen, John McGreevy for insightful comments. The authors acknowledge the support provided by the US Department of Energy (DOE) under cooperative research agreement DE-SC0009919.

\bibliographystyle{apsrev}
\bibliography{refs}

\appendix
\section{Diagonalizable and finite dimensional dilatation generator}
\label{diagonalizableD}
In this appendix, we re-derive some of the results in Sec.~\ref{main} under  the stronger assumption that  the matrix
$\mathbf{\Delta}$ is both  diagonal and finite dimensional. This discussion is intended for clarity,
since it is less abstract than the one presented in the main text. 

We recall that
\begin{align}\label{genDapp}
[D,\tilde{\Phi}(\vec{x}=\vec{0},t=0)]&=\imath\mathcal{D}\tilde{\Phi}(\vec{x}=\vec{0},t=0)
\end{align}
and $\mathcal{D}$ is renamed as $\mathbf{\Delta}$ in the finite dimensional case. 

To warm up, we show that  both $\mathcal{D}$ and $\mathcal{N}$  are hermitian only if
$z=2$ or $\mathcal{N}=0$.  From  $[D,N]=\imath (2-z)N$, it follows that
\begin{equation}
[\mathcal{D},\mathcal{N}]=(2-z)\mathcal{N}\,.
\end{equation}
Since $\mathcal{D}$ and $\mathcal{N}$  are assumed hermitian, $[\mathcal{D},\mathcal{N}]^{\dagger}=-[\mathcal{D},\mathcal{N}]=-(2-z)\mathcal{N}$. Hence
\begin{equation*}
\nonumber -(2-z)\mathcal{N}=[\mathcal{D},\mathcal{N}]^{\dagger}=(2-z)\mathcal{N}^{\dagger}=(2-z)\mathcal{N}\,,
\end{equation*}
which can only hold for $z=2$ or $\mathcal{N}=0$. If one assumes $\mathcal{N}\ne0$ for
some field, then $z=2$. One can have $z\ne2$ and hermiticity if $\mathcal{N}=0$ for all
fields. In this case the generator $N$ is superfluous, and one can extend the algebra by
including the generator of special conformal transformations.\footnote{There are indeed examples of
  $z\neq2$ theories without particle number symmetry; see, for example, Refs.~\cite{Hofman:2014loa,Jensen:2017tnb,Stichel:2010ue}.} Below we assume $N$ does not
identically vanish. Similarly, both $\mathcal{D}$ and $\mathcal{K}$  are hermitian only if
$z=1$ or $\mathcal{K}=0$.

Now we consider the finite dimensional case where $\mathbf{\Delta}$ is diagonal. Alternatively, one
can consider the case that $\mathbf{\Delta}$ is hermitian (and therefore,  as just proved,
$\mathbb{N}$ is not hermitian). In the finite dimensional, hermitian case, one can always choose to
diagonalize $\mathbf{\Delta}$.  Since $\mathbf{\Delta}$ is diagonal,  $[\mathbf{\Delta},\mathbb{N}]=(2-z)\mathbb{N}$ implies that
$\left(\mathbf{\Delta}_{\alpha\alpha}-\mathbf{\Delta}_{\beta\beta}+z-2\right)\mathbb{N}_{\alpha\beta}=0$
(no summation over indices $\alpha,\beta$ is implicit), which, in turn, for $z\neq2$ implies that
$\mathbb{N}_{\alpha\alpha}=0$  and  at least one
of 
$\mathbb{N}_{\alpha\beta}$  and $\mathbb{N}_{\beta\alpha}$ vanish. This implies that 
$\mathbb{N} $ is nilpotent,
\begin{align}
\mathbb{N}^{M}=0\,,
\end{align} 
for some integer $M$ no larger than the dimension of the representation.  One can show this, without
loss of generality, by arranging the components of the fields $\tilde{\mathbf{\Phi}}_\alpha$ so that
$\mathbb{N}$ is an upper triangular matrix. This result will play a pivotal role below.

Similarly, we assume that the field $\tilde{\mathbf{\Phi}}(\vec{0},0)$ has the following commutation relation:
\begin{align}
\label{gen1}
[K_i,\tilde{\mathbf{\Phi}}(\vec{x}=\vec{0},t=0)]&=\mathcal{K}_i\tilde{\mathbf{\Phi}}(\vec{x}=\vec{0},t=0)\,.
\end{align}
For a finite dimensional case, we denote $\mathcal{K}_i$ by $\mathbb{K}_i$. By the same argument as
above, one can show that either $z=1$ or $\mathbb{K}_i=0$ or $\mathbb{K}_i$ is nilpotent. Thus we have
\begin{align}
\mathbb{K}_{i}^{L_{i}}=0\,,
\end{align} 
for some integer $L_{i}$ no larger than the dimension of the representation. One can consider the operator
\begin{align}
\mathbf{\Phi}\equiv \prod_{i=1}^{i=d-1} K_{i}^{L_i-1} (\tilde{\mathbf{\Phi}})\,,
\end{align}
where for any operator $A$ and $B$, the action of the operator on the field is defined via
\begin{align}
A (\tilde{\mathbf{\Phi}})& \equiv [A,\tilde{\mathbf{\Phi}}]\,,\\
BA(\tilde{\mathbf{\Phi}}) &\equiv B \big(A(\tilde{\mathbf{\Phi}}) \big)\,.
\end{align}

It can be easily verified that
\begin{align}
\label{prime}\left[K_i,\mathbf{\Phi}(\vec{x}=\vec{0},t=0)\right]&=0\,,\\
\left[D,\mathbf{\Phi}(\vec{x}=\vec{0},t=0)\right]&=\imath\mathbb{\Delta}^{\prime}\mathbf{\Phi}(\vec{x}=\vec{0},t=0)\,,\\
\left[N,\mathbf{\Phi}(\vec{x},t)\right]&=\mathbb{N}\mathbf{\Phi}(\vec{x},t)\,,
\end{align}
where $\mathbb{\Delta}^{\prime}=\left(\mathbf{\Delta}-(z-1)\sum_i(L_i-1)\right)$. We call `primary
operators' those that satisfy \eqref{prime}. One could have considered operators obtained from these
by analogous operations as above i.e. operators of the form $[N^{M-1},\mathbf{\Phi}]$, but that would not suffice to reveal the problems associated with finite dimensional representations. 

Consider the two point correlator of primary operators in such a realization of the algebra, $G_{\alpha\beta}\equiv \langle 0|\mathbf{\Phi}_{\alpha}(x,t)\mathbf{\Phi}_{\beta}(0,0)|0\rangle$. Using
Eqs.~\eqref{eq:canonicalHP}, the commutator in~\eqref{prime}  translates to
\begin{align}
[K_i, \mathbf{\Phi}]&= \left(-\imath t\partial_{i} \mathbf{I}+ x_i\mathbb{N}\right)\mathbf{\Phi}\,.
\end{align}
Galilean boost invariance of the vacuum, $K_i|0\rangle =\langle 0|K_i=0$, then
gives
\begin{multline}
\nonumber \langle 0|\left[K,\mathbf{\Phi}_\alpha(x,t)\mathbf{\Phi}_\beta(0,0)\right]|0\rangle = 0\\
\Rightarrow \left(-\imath t\partial_{i} \delta_{\alpha\sigma}+ x_i\mathbb{N}_{\alpha\sigma}\right)G_{\sigma\beta}=0\,.
\end{multline}
The solution to the above differential equation  is given by
\begin{equation}\label{correlator}
 G_{\alpha\beta}=
\!\!\left[e^{-\imath\frac{|\vec{x}|^2}{2t}\mathbb{N}}\right]_{\alpha\gamma}\!\!\!\!C_{\gamma\beta}(t)
= \!\!\sum_{\ell=0}^{\substack{M-1}}\frac{1}{\ell!}\!\! \left(-\imath\frac{|\vec{x}|^2}{2t}\right)^{\!\!\ell} \!\!(\mathbb{N}^{\ell}C(t))_{\alpha\beta}
\end{equation}
where $|\vec{x}|^{2}=\sum_i(x^{i})^2$, $C$ is an as yet undetermined matrix function of $t$ alone;
the scaling symmetry implies that $C_{\alpha\beta}$ has a power law dependence on $t$. The above
correlator~\eqref{correlator} is consistent with the one given in \eqref{maincorrelator} with
$\mathbb{N}_1=\mathbb{K}_1=\mathbb{K}_2=0$. The exponential becomes a finite degree polynomial
because $\mathbb{N}$ is nilpotent, and this is very specific to $z\neq 2$ theories. As explained
above, the correlators are badly behaved: polynomial rather than exponential dependence on
$|\vec{x}|$ leads to growth with spatial separation (and hence, cluster decomposition fails).  In
contrast, for $z=2$ the matrix $\mathbb{N}$ is diagonal and there is no truncation of the expansion
of the exponential. An additional constraint on the correlator follows from requiring that
$\langle 0| \left[N,\mathbf{\Phi}_{\alpha}(x,t)\mathbf{\Phi}_{\beta}(0,0)\right]|0\rangle=0$, which
implies $NG+GN^{T}=0$.

Consider next $G^{\prime}_{\alpha\beta}=\langle 0|\mathbf{\Phi}_{\alpha}(x,t)\mathbf{\Phi}^{\dagger}_{\beta}(0,0)|0\rangle$. This is similarly  given by  
\begin{align}
G^{\prime}_{\alpha\beta}=\left[\exp\left(-\imath\frac{|\vec{x}|^2}{2t}\mathbb{N}\right)C'(t)\right]_{\alpha\beta}
\end{align}
for some undetermined matrix $C^{\prime}$ such that $C^{\prime}_{\alpha\beta}$ is a function of $t$ alone. Invariance under $N$ implies that $NG^{\prime}-G^{\prime}N^{\dagger}=0$. Notice that the condition on $G'$ is different from that on $G$; one may have non-trivial solutions
to one but not the other.  For example, one can consider the two
component field, $\mathbf{\Phi}_{\alpha=1,2}$ characterized by:
\begin{gather*}
\mathbb{N}=
\begin{bmatrix}
0 & 1\\
0 & 0
\end{bmatrix}\,, \quad C^{\prime}=g(t)\begin{bmatrix}
0 & 1\\
1 & 0
\end{bmatrix}\,,\ g(t)={t^{-({\Delta}^{}_{11}+\Delta^{}_{22})/z}}\\[1ex]
\mathbb{\Delta}=\begin{bmatrix}
\Delta_{11} & 0\\
0 & \Delta_{22}
\end{bmatrix}\,,\quad \Delta^{}_{22}-\Delta^{}_{11}=(2-z)\,,\\[1ex]
G^{\prime}={t^{-({\Delta}^{}_{11}+\Delta^{}_{22})/z}}\begin{bmatrix}
1 & -\imath \frac{|\vec{x}|^2}{2t}\\
0 & 1
\end{bmatrix}\begin{bmatrix}
0 & 1\\
1 & 0
\end{bmatrix}\\
\qquad\qquad\qquad\qquad={t^{-({\Delta}^{}_{11}+\Delta^{}_{22})/z}}\begin{bmatrix}
-\imath \frac{|\vec{x}|^2}{2t} & 1\\
1 & 0
\end{bmatrix}\,.
\end{gather*}
Note that for this example $G_{\alpha\beta}=0$ so consideration of the long distance behavior of
this correlator alone does not, by itself, suggest the theory is ill-behaved. 
\end{document}